# Validity of threshold-crossing analysis of symbolic dynamics from chaotic time series

Erik M. Bollt[1], Theodore Stanford[2], Ying-Cheng Lai[3], Karol Życzkowski[4,5]

1,2: Mathematics Department, 572 Holloway Rd, U.S. Naval Academy, Annapolis, MD 21402-5002.
3: Departments of Physics and Astronomy and of Mathematics, The University of Kansas, Lawrence, Kansas 66045.
4: Instytut Fizyki im. Smoluchowskiego, Uniwersytet Jagielloński, ul. Reymonta 4, 30-059 Kraków, Poland.
5: Centrum Fizyki Teoretycznej PAN, al. Lotników 32/46, 02-668 Warszawa, Poland.

PACS: 05.45.-a

**Abstract**

A practical and popular technique to extract the symbolic dynamics from experimentally measured chaotic time series is the threshold-crossing method, by which an arbitrary partition is utilized for determining the symbols. We address to what extent the symbolic dynamics so obtained can faithfully represent the phase-space dynamics. Our principal result is that such practice leads to a severe misrepresentation of the dynamical system. The measured topological entropy is a Devil's staircase-like, but surprisingly nonmonotone, function of a parameter characterizing the amount of misplacement of the partition.

Symbolic dynamics is a fundamental tool available to describe complicated time evolution of a chaotic dynamical system, the Smale horseshoe [1] being the most famous proto-type. Symbolic dynamics also provides a natural link between chaotic dynamics and information theory [2], on which the recent idea of utilizing chaotic systems to encode digital information, or communicating with chaos, is based [3]. A good symbolic dynamical representation requires that a one-to-one correspondence be established to the phase-space dynamics; the partition that defines distinct symbols has to be *generating* [4]. Specification of the generating partition for chaotic systems in general is, however, a challenging problem [5, 6, 7, 8] which is still open for most dynamical systems.

On the experimental side, there appears an increasing interest in chaotic symbolic dynamics as well [9, 10]. A common practice is to apply the threshold-crossing method, *i.e.*, to define a rather arbitrary partition, so that distinct symbols can be defined from measured time series [9]. There are two reasons for the popularity of the threshold-crossing method: (1) it is extremely difficult to locate the generating partition from chaotic data, and (2) threshold-crossing is a physically intuitive and natural idea. Consider, for instance, a time series of temperature $T(t)$ recorded from a turbulent flow. By replacing the real-valued data with symbolic data relative to some threshold $T_c$, say a **0** if $T(t) < T_c$ and a **1** if $T(t) > T_c$, the problem of data analysis can be simplified. A well chosen partition is clearly important: for instance, $T_c$ cannot be outside the range of $T(t)$ because, otherwise, the symbolic sequence will be trivial and carry no information about the underlying dynamics. It is thus of paramount interest, from both the theoretical and experimental points of view, to understand how misplaced partitions affect the goodness of the symbolic dynamics such as the amount of information that can be extracted from the data.

In this Letter, we investigate the consequence of misplaced partitions in chaotic systems. Specifically, we address how the topological entropy, perhaps one of the most important dynamical invariants that one intends to compute from symbolic dynamics, behaves as a parameter, $d$, which characterizes the amount of misplaced partition, is changed. We find that the topological entropy as a function of $d$ to be devil's staircase-like, but surprisingly nonmonotone. We establish our results by performing numerical computations for one- and two-dimensional maps, and by rigorous analyses for the tent map which is a good topological model for one-dimensional one-hump maps. The main implication of our results is that the threshold-crossing technique typically yields misleading conclusions about the dynamics generating the data, and therefore one should be extremely cautious when attempting to understand the underlying system from a misrepresented symbolic dynamics. A previous study on the role of noise on the symbolic dynamics [18] concludes that entropy increases according to a power law with noise volume, and that fine structures of invariant measure blur at the noise threshold. Similar topological entropy figures have appeared previously [18], but our combinatorial and topological explanation of the phenomenon and application are new.

We begin by studying the tent map: $f : [0, 1] \to [0, 1], x \to 1 - 2|x - 1/2|$ for which the generating partition for symbolic dynamics is the critical point $x_c = 1/2$. The dynamics of $f$ is semi-conjugate to the

Bernoulli full 2-shift, meaning that there are no forbidden binary symbolic sequences, by the surjection $h : [0,1] \to \Sigma_2^{\{\mathbf{0,1}\}}$. This gives a unique symbolic itinerary $\sigma = \sigma_0.\sigma_1\sigma_2\ldots$ for each $x \in [0,1]$, where $\sigma_i(x) = \mathbf{0(1)}$ if $f^i(x) < x_c(> x_c)$ for $i \geq 0$, and $\Sigma_2^{\{\mathbf{0,1}\}}$ is the semi-infinite fullshift on two symbols: $\mathbf{0}$ and $\mathbf{1}$ [11, 12]. The topological entropy of $\Sigma_2^{\{\mathbf{0,1}\}}$ is $\ln 2$. Now misplace the partition at $p = x_c + d$, where $d \in [-1/2, 1/2]$ is the misplacement parameter. In this case, the symbolic sequence corresponding to a point $x \in [0,1]$ becomes: $\phi = \phi_0.\phi_1\phi_2\ldots$, where $\phi_i(x) = \mathbf{a(b)}$ if $f^i(x) < p(> p)$, as shown in Fig. 1. The shift so obtained: $\Sigma_2^{\{\mathbf{a,b}\}}$, will no longer be a full shift because, as we will argue later, not every binary symbolic sequence is possible. Thus, $\Sigma_2^{\{\mathbf{a,b}\}}$ will be a subshift on two symbols $\mathbf{a}$ and $\mathbf{b}$ when $d \neq 0$ ($p \neq x_c$). The topological entropy of the subshift $\Sigma_2^{\{\mathbf{a,b}\}}$, denoted by $h_T(d)$, will typically be less than $h_T(0) = \ln 2$. Numerically, $h_T(d)$ can be computed by using the formula [12]:

$$h_T(d) = \lim_{n \to \infty} sup \frac{\ln N_n}{n}, \quad (1)$$

where $N_n \leq 2^n$ is the number of $(\mathbf{a}, \mathbf{b})$ binary sequences (words) of length $n$. In our computation, we choose 1024 values of $d$ uniformly in the interval $[-1/2, 1/2]$ and for each value of $d$, we count $N_n$ in the range $4 \leq n \leq 18$ from a trajectory of $2^{20}$ points generated by the tent map. The slopes of the plots of $\ln N_n$ versus $n$ gives approximate values of $h_T$. Figure 2(a) shows $h_T(d)$ versus $d$ for the tent map, where we observe a complicated, devil's staircase-like, but clearly nonmonotone behavior. For $d = 0$, we have $h_T(0) \approx \ln 2$, as expected. For $d = -1/2$ (1/2), from Fig. 1, we see that the grammar forbids the letter $\mathbf{a}$ ($\mathbf{b}$) and, hence, $\Sigma_2^{\{\mathbf{a,b}\}}(-1/2)$ [$\Sigma_2^{\{\mathbf{a,b}\}}(1/2)$] has only one sequence: $\phi = \mathbf{b.b\bar{b}}$ ($\phi = \mathbf{a.a\bar{a}}$). Hence, $h_T(\pm 1/2) = 0$. Intuitively, one might imagine that due to the two-to-one nature of the tent map, placing the partition at any $d \neq 0$ yields two points: $x, \bar{x} \in (x_c - |d|, x_c + |d|)$, which are symmetrically placed relative to $x_c$, to have the same future orbit and identical first bit of the symbolic sequence. Hence, the two points are indistinguishable. Such a loss of symmetric words may lead one to think that $h_T(d)$ should be a monotonically decreasing function of $d$. As we will argue rigorously later, such an intuition is incorrect.

Can a similar behavior in the topological entropy occur in more complicated and nonhyperbolic systems? Consider, for instance, the Hénon map: $(x, y) \to (1.4 - x^2 + 0.3y, x)$. Due to nonhyperbolicity, the generating partition has been conjectured to be a complicated zig-zag curve connecting all primary tangency points between the stable and the unstable manifolds [5, 13]. Precisely locating the partition curve is highly nontrivial, so the idea of threshold-crossing becomes more tempting. Suppose we simply set the partition at $y = c$ and define, for each measurement $y_i$ from the time series $\{y_i\}_{i=0}^N$, the following $2n$-bit word: $\omega_{2n} = \sigma_{-n} \ldots \sigma_{-1}.\sigma_0 \ldots \sigma_{n-1}$, where $n \leq i \leq N - n$ and the symbols are chosen according to the rule: $\sigma_j = \mathbf{a}(\mathbf{b})$ if $y_{i+j} < c (> c)$. By numerically counting the number of distinct $2n$-bit words, we obtain Fig. 3, plot of $h_T(c)$ versus the misplaced partition parameter $c$. Features similar to those in Fig. 2(a) are observed. Computations using experimental time series such as the Belousov-Zhabotinskii reaction data [14] reveal a similar behavior depicted in Figure 5.

Our experimental and numerical results are justified by the following rigorous analysis [17]. Consider a graphic presentation of the Bernoulli shift map in symbolic space. For instance, the fullshift $\Sigma_2^{\{\mathbf{0},\mathbf{1}\}}$ can be naturally presented by a directed graph describing all possible transitions between the $2^n$ words of **0**'s and **1**'s, as shown in Fig. 4 for $n = 4$. The sixteen 4-bit words are arranged from left to right by the Gray-code order: $\mathbf{0000} \prec \mathbf{0001} \prec \mathbf{0010} \prec \mathbf{0110} \prec \ldots \prec \mathbf{1011} \prec \mathbf{1001} \prec \mathbf{1000}$, so that corresponding intervals in the phase space are monotonically ordered, according to the kneading theory [15]. In Fig. 4, several cases are shown for which the partition is misplaced at dyadic positions: $d = q/2^n \equiv q/r$, where $d \in [0,1]$ and $q, r, n$ are integers. A *sophic shift* $X$ is defined to be [16] a shift space generated by all possible walks through a labeled graph $\mathcal{G} = (G, L)$, where each edge carries a label index $L$, and a particular element of $X$ is defined by the labels of the set of edges followed during a particular infinite walk. A sophic shift is *right resolving* if each vertex has all of its exiting edges labeled uniquely [16]. Figure 4 is, for example, a right resolving presentation of a sophic shift that is conjugate to the full 2-shift.

Consider a misplacement which occurs at a dyadic $d$, then the partition of the tent map can be described in the symbolic space by relabeling the set of edges of an appropriate graph $(G, L)$ presentation of the fullshift. Placing the partition at any such $d$ is equivalent to relabeling the set of edges in the graphic presentation, to be either **a** or **b**, according to the Gray-code order. In particular, if $d \preceq v_i$, where $v_i$ ($i = 1, \ldots, 2^n$) is the $i^{th}$ $n$-bit word labeling a vertex, then all edges pointed *into* that vertex are relabeled as an **a**, and otherwise a **b**. Such a relabeling can be considered to be a factor code $\phi_d$ that projects, surjectively, the fullshift $\Sigma_2^{\{0,1\}}$ into the subshift $\Sigma_2^{\{a,b\}}(d)$. Each relabeled graph then defines a new subshift with grammar (because there are now forbidden words) by considering all walks through the graph and all resulting infinite sequences of **a**'s and **b**'s.

Consider the following examples. If $n = 4$ and $d = -1/8$, it can be shown [17] that the relabeled word **abbba** is allowed which is generated by the path: $\overbrace{0111}^{a} \to \overbrace{1110}^{b} \to \overbrace{1100}^{b} \to \overbrace{1001}^{b} \to \overbrace{0010}^{a}$. This is in fact the only such labeled path through the graph, as shown in Fig. 4. If the partition is further misplaced to $d = -3/16$, then that path becomes relabeled: **bbbba**, since we have, by relabeling the vertex **0111** to a **b**, the following: $\overbrace{0111}^{b} \to \overbrace{1110}^{b} \to \overbrace{1100}^{b} \to \overbrace{1001}^{b} \to \overbrace{0010}^{a}$. For $d = -1/8$, it is easy to show [17], by considering the short list of all possible 5-bit paths out of all 16 vertices, that this is the only path that gives **abbba** and, hence, when $d = -3/16$, **abbba** can label no other paths, and as such the word becomes forbidden. A key point to notice is that the word **abbba** will not be forbidden when the partition is even further misplaced to $d = -1/4$: $\overbrace{0011}^{a} \to \overbrace{0110}^{b} \to \overbrace{1100}^{b} \to \overbrace{1000}^{b} \to \overbrace{0000}^{a}$, which was formerly named **aabba**. This example gives some intuitive illustratation that $h_T(d)$ as a function of $d$ can be nonmonotone, because words that disappear can reappear, by different paths, as $|d|$ is increased.

We now explain a procedure to rigorously compute the entropy function $h_T(d)$. First we recall two important theorems from Ref. [16]: 1) If $X$ is a sophic shift and $\mathcal{G} = (G, L)$ is a right resolving presentation of $X$, then $h_T(X) = \ln \rho[A(\mathcal{G})]$, where $\rho[A(\mathcal{G})]$ is the spectral radius of the transition matrix $A$ corresponding to $\mathcal{G}$. 2) Every sophic shift has a right resolving presentation. The following proof is equivalent to the subset construction in Ref. [16], which allows us to prove that a right resolving presentation of these misplaced sophic shifts $\Sigma_2^{\{a,b\}}(d)$ can be so constructed for any dyadic misplacement, $d = q/2^n$. Then $\Sigma_2^{\{a,b\}}(d)$ can be presented by relabeling the $N = 2^n$ nodes in the graphic presentation of the fullshift $\Sigma_2^{\{0,1\}}$, as we have discussed in the preceding paragraphs. Define the graph that generates $X_{\mathcal{G}} \equiv \Sigma_2^{\{a,b\}}(d)$ to be $\mathcal{G} = (G, L)$. Now index each of the $N$ vertices of $G$: $G_i$ ($i = 1, \ldots 2^n$) and define a new set of vertices: $H = \{H\}_{j=1}^{2^n}$, where each $H_j$ defines specific and unique on/off states for each vertex.

For convenience, we label each $H_j$ by $N$ $X$'s and $O$'s, each's giving an on/off state for each $G_i$. In this notation, if only $G_i$ is "turned on," then $H_j = OO\ldots OXO\ldots O$ has an $X$ (on) only in the $i^{th}$ position. Considering all on/off states of $\{G_i\}_{i=1}^N$ requires $2^N$ vertices $H_j$. The next step is to define edges $E$ to go with the vertices $H$. Each $H_j$ labels on/off positions of $N$ vertices of $G$. Consider all the vertices $G_k$ that can follow $G_i$ such that each $G_k$ has the label **a** to be "turned on" and all other $G_k$'s to be "turned off." If there is no transition from $G_i$'s to an **a**-labeled $G_k$, then no edge will be defined. Thus, for each nonempty transition, a vertex $H_l$ is defined to follow $H_j$ with an **a**. Similarly, we can define **b** edges from $H_j$. Following these rules, we construct a graph $\mathcal{H} = (H, E)$ that generates a sophic shift $\mathcal{X}_\mathcal{H}$. We advance the following two propositions [17]:

<u>Proposition 1</u>: $\mathcal{X}_\mathcal{H}$ is conjugate to $\Sigma_2^{\{\mathbf{a},\mathbf{b}\}}(d)$ and, hence, $h_T(\mathcal{X}_\mathcal{H}) = h_T[\Sigma_2^{\{\mathbf{a},\mathbf{b}\}}(d)]$.

<u>Proposition 2</u>: $\mathcal{H} = (H, E)$ is a right resolving presentation of $\Sigma_2^{\{\mathbf{a},\mathbf{b}\}}(d)$.

Our algorithm to computer $h_t(d)$ consists of: 1) for each $d = q/2^n$, find the corresponding transition graph $\mathcal{G}$, 2) generate a right resolving presentation $\mathcal{H}$, and its transition matrix $A(\mathcal{H})$, 3) compute topological entropy by the largest eigenvalue of $A(\mathcal{H})$. The result is shown in 2(b). We consider Fig. 2(b) to be essentially exact, although the eigenvalues for such large transition matrices (e.g., we have used $n = 16$) need to be computed numerically. It should be mentioned that while the above algorithm is guaranteed to produce a finite, but large $2^n$-vertex graph $\mathcal{H}$, given an initial $n$-vertex graph $\mathcal{G}$, we can show that the problem is numerically tractable since much smaller irreducible subcomponents are expected [17].

In summary, we have presented numerical and rigorous results concerning the behavior of the topological entropy when the partition for symbolic dynamics is misplaced, which has been a common practice utilized to extract the symbolic dynamics from experimentally measured chaotic data. Our principal result is that the entropy can be a nonmonotone and devil's staircase-like function of the misplacement parameter. As such, the consequence of a misplaced partition can be severe, including significantly reduced topological entropies and a high degree of nonuniqueness which we will rigorously elaborate else-where [17]. We wish to convey the message in this Letter that interpreting any results obtained from threshold-crossing type of analysis should be exercised with extreme caution.

EMB was supported by NSF under Grant No. DMS-9704639. TBS was supported by N.A.R.C. YCL was supported by AFOSR under Grant No. F49620-98-1-0400 and by NSF under Grant No. PHY-9722156. K. Ż. is thankful to USNA for hospitality. The authors are thankful to Eric Kostelich who provided the BZ data [14].

# References

[1] S. Smale, Bull. Ame. Math. Soc. **73**, 747 (1967).

[2] C. E. Shannon and W. Weaver, *The Mathematical Theory of Communication* (Univ. of Illinois Press, 1964).

[3] S. Hayes, C. Grebogi, and E. Ott, Phys. Rev. Lett. **70**, 3031 (1993); E. Rosa, S. Hayes, and C. Grebogi, Phys. Rev. Lett. **78**, 1247 (1997); E. Bollt, M. Dolnik, Phys. Rev. E **55**, 6404 (1997); E. Bollt, Y.-C. Lai, C. Grebogi, Phys. Rev. Lett. **79**, 3787 (1997); E. Bollt and Y.-C. Lai, Phys. Rev. E **58**, 1724 (1998).

[4] D. J. Rudolph, *Fundamentals of Measurable Dynamics, Ergodic Theory on Lebesgue Spaces* (Clarendon Press, Oxford, 1990).

[5] P. Grassberger, H. Kantz, and U. Moenig, J. Phys. A: Math. Gen. **22**, 5217 (1989).

[6] F. Christiansen and A. Politi, Phys. Rev. E **51**, R3811 (1995), Nonlinearity **9**, 1623 (1996).

[7] G. Boulant, M. Lefranc, S. Bielawski, D. Derozier, Phys. Rev. E **55** 5082 (1997).

[8] R. L. Davidchack, Y.-C. Lai, E. M. Bollt, and M. Dhamala, Phys. Rev. E **61**, 1353 (2000).

[9] J. Kurths, *et. al.*, Chaos **5**, 88 (1995); M. Lehrman and A. B. Rechester, Phys. Rev. Lett. **78**, 1 (1997); C. S. Daw, *et. al.*, Phys. Rev. E **57**, 2811 (1998); R. Engbert, C. Scheffczyk, R. Krampe, J. Kurths,

and R. Kliegl, pp. 271-282 in *Nonlinear Time Series Analysis of Physiological Data* (Springer-Verlag, Heidelbert, 1998).

[10] K. Mischaikow, *et. al.*, Phys. Rev. Lett. **82**, 1144 (1999).

[11] R. L. Devaney, *An Introduction to Chaotic Dynamical Systems* (Addison-Wesley, Reading MA, 1989).

[12] C. Robinson, *Dynamical Systems: Stability, Symbolic Dynamics, and Chaos* (CRC Press, Ann Arbor, 1995).

[13] P. Cvitanovic, G. Gunaratne, and I. Procaccia, Phys. Rev. A **38**, 1503 (1988).

[14] R. H. Simoyi, A. Wolf, and H. L. Swinney, Phys. Rev. Lett. **49**, 245 (1982).

[15] J. Milnor and W. Thurston, *On Iterated Maps of the Interval (I and II)* (Princeton, 1977).

[16] D. Lind and B. Marcus, *An Introduction to Symbolic Dynamics and Coding* (Cambridge Univ. Press, New York, 1995).

[17] The details of proofs, together with examples illustrating the construction of the sophic shift for symbolic dynamics with misplaced partition, will be published in a future paper with explanation of nonmonotonicity of the topological entropy function by masking functions, issues of nonuniqueness of symbolic sequences, and many other details [E. M. Bollt, T. Stanford, Y.-C. Lai, and K. Życzkowski, preprint (2000)].

[18] J.P.Crutchfield, N.H. Packard, Int. J. Th. Phys. Vol. 21, Nos 6/7, 1982; J.P. Crutchfield, N.H. Packard, Physica 7D, 201-223 (1983); C. M. Glenn, S. Hayes, "Targeting Regions of Chaotic Attractors USing Small Perturbation Control of Symbolic Dynamics," ARL-TR-903, May 1996.

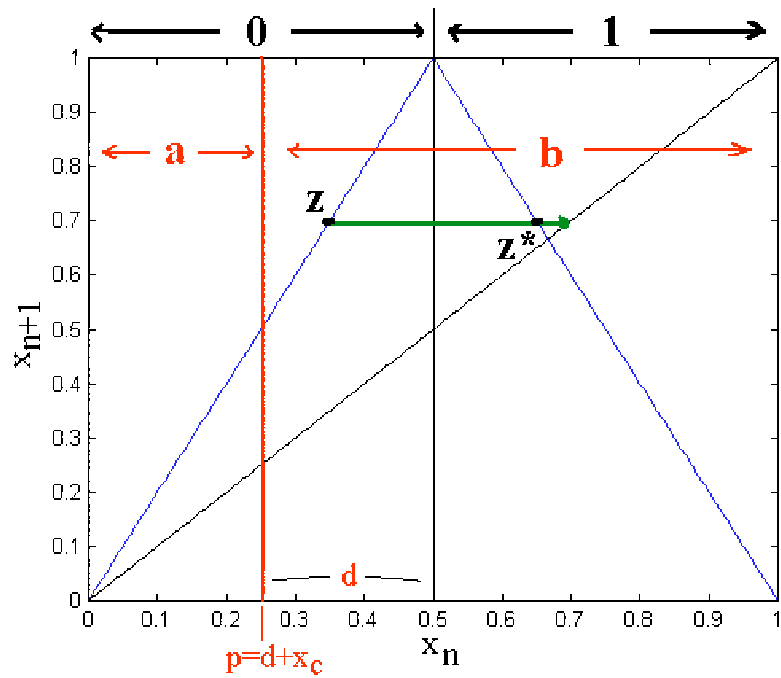

Figure 1: Tent map and a misplaced partition at $x = p$.

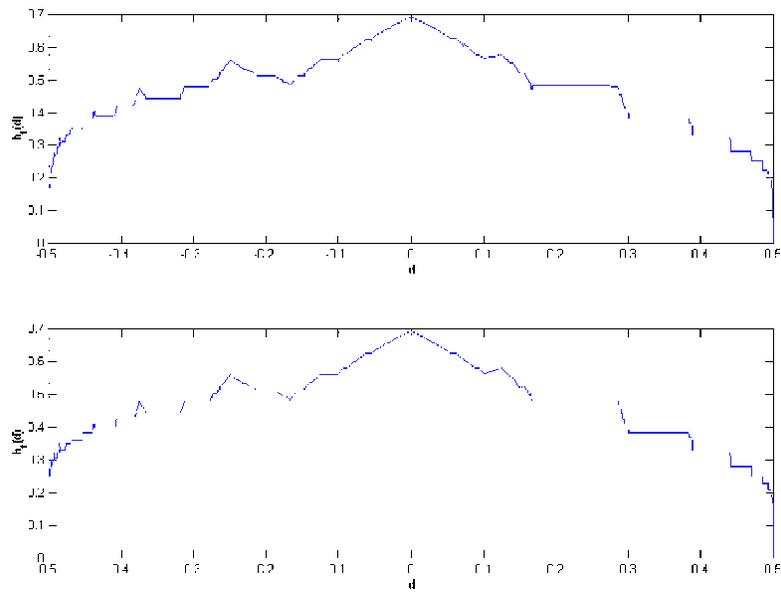

Figure 2: For the tent map: (a) numerically computed $h_t(d)$ function by following sequences of a chaotic orbit; and (b) exactly computed entropy function $h_T(d)$.

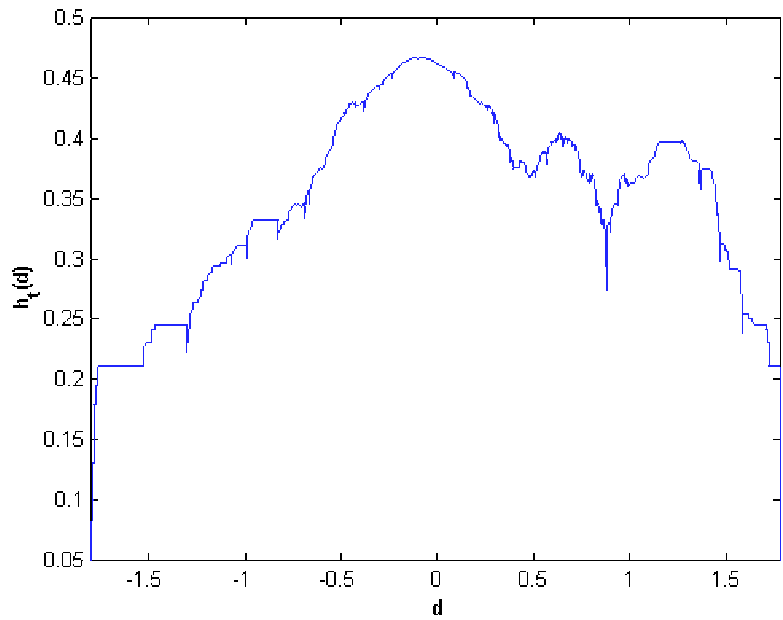

Figure 3: For the standard Hénon map, the topological entropy $h_T$ versus $c$, the parameter characterizing the misplaced partition at $y = c$. Features similar to those in Fig. 2 are seen.

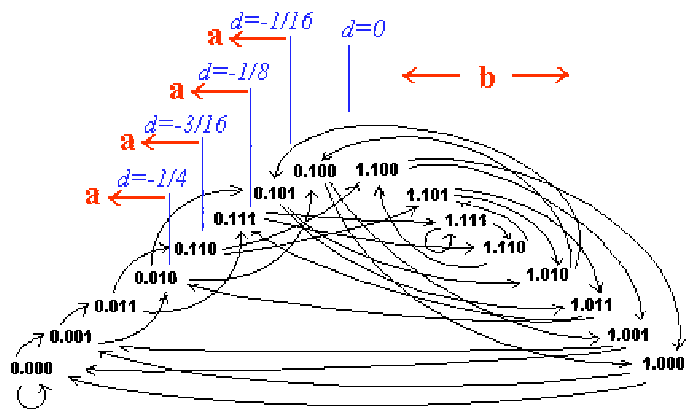

Figure 4: Graphic presentation for the Bernoulli fullshift and some dyadic misplacements

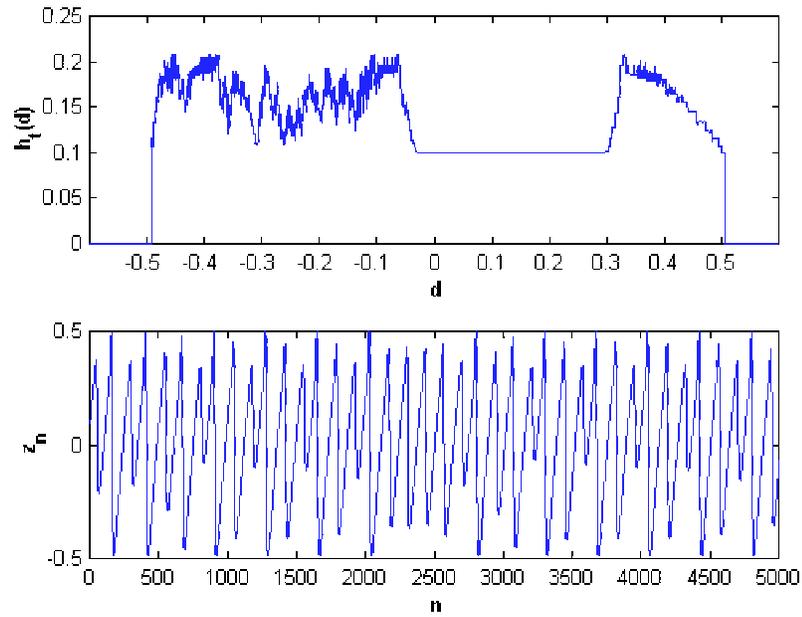

Figure 5: (a) Topological entropy versus partition placement, $h_t(d)$ for experimentally collected chemical reaction (BZ) data. (b) Time-series of the data.